\documentstyle[twocolumn,floats,aps,epsf]{revtex}
\begin{document}

\twocolumn[\hsize\textwidth\columnwidth\hsize\csname@twocolumnfalse%
\endcsname

\title{The role of intermediate layers in the $c$ axis conductivity of
layered superconductors}
\author{W. A. Atkinson}
\address{Department of Physics, Indiana University,
Swain Hall W.\ 117, Bloomington IN 47405}
\author{W. C. Wu, and J. P. Carbotte}
\address{Department of Physics and Astronomy, McMaster University,
Hamilton, Ontario, Canada L8S 4M1}
\date{\today}
\maketitle
\draft
\begin{abstract}
A simplified model of $c$ axis transport in the high $T_c$ superconductors
is presented.  Expressions are found for the $c$ axis optical conductivity,
the d.c.\ resistivity, and the $c$ axis penetration depth.  Within the
framework of this model, the pseudogap in the optical conductivity
arises naturally as a result of the layered band structure of the 
high $T_c$ materials.  We discuss the occurence
of the pseudogap in terms of three parameters:  a band gap
$\Delta_{\mbox{\scriptsize ps}}$, 
a temperature dependent scattering rate $\Gamma(T)$,
and the strength of the interlayer coupling $t_\perp$.  
We are also able to find analytic expressions for the d.c.\ conductivity
and the low temperature penetration depth in terms of these three parameters.
This work is an attempt to present a simple, unified picture of $c$
axis properties in the high $T_c$ cuprates.
\end{abstract}

]
\narrowtext
Recently, a study was made\cite{atkinsonI} of the effect of band
structure in YBCO on the real part of the $c$ axis optical
conductivity $\sigma_c(\omega)$.  It was found that interband
transitions between the plane and chain derived bands make a large 
contribution to the $c$ axis conductivity, while having almost no effect
on the in-plane conductivities.  The interband term results
in a $c$ axis conductivity which is qualitatively different from the
in-plane conductivity, and which seems to explain many
of the features seen in experiments.

In Ref.~\onlinecite{atkinsonI}, the YBCO crystal is treated as a
system of two-dimensional plane, and one-dimensional chain layers,
coupled through a hopping matrix element $t_\perp$.  While the model
is grossly oversimplified, it is still too complicated for anything
more than a numerical treatment,  making it difficult to understand
explicitely the role of the various model parameters.

In this work, a much simpler model is presented for which it is
possible to find analytical expressions for the optical conductivity,
the d.c.\ resistivity and the low-temperature penetration depth.  We
will identify three important parameters which influence the
qualitative structure of the conductivity: $t_\perp$ is the strength
of the coupling between the layers, $\Delta_{\mbox{\scriptsize ps}}$
is the minimum energy difference between the bands and $\Gamma(T)$ is
a temperature dependent scattering rate.

Let us first address the issue of when one can expect interband
transitions to be important.  Consider a system made up of two types
of layer, stacked in alternating fashion along the $c$ axis.  Let the
dispersions for the isolated layers be $\xi_1(k_x,k_y)$ and
$\xi_2(k_x,k_y)$.  The layers are coupled via single electron hopping,
with a matrix element $t_\perp$.  It is assumed that the first layer,
with dispersion $\xi_1$ represents a CuO$_2$ layer.  The second layer,
with dispersion $\xi_2$, is the ``intermediate layer''.  In the case
of YBCO, the intermediate layer is the CuO chain layer.  In other
materials, the choice of intermediate layer is not clear and, in fact,
there will generally be several intermediate layers.  Even in the
simple case of LSCO, an electron must find some path through two
intermediate LaO layers when travelling between adjacent CuO$_2$
planes.  

The Hamiltonian has the form
\[
H = \sum_{\bf k} \left[ \begin{array}{cc} 
c_{1{\bf k}}^\dagger & c_{2{\bf k}}^\dagger
\end{array} \right ] 
\, h({\bf k}) \,
\left[ \begin{array}{c} 
c_{1{\bf k}} \\ c_{2{\bf k}}
\end{array} \right ],
\]
where
\[
h({\bf k}) = 
\left[ \begin{array}{cc} 
\xi_1(k_x,k_y) & t(k_z) \\ t^\ast(k_z) & \xi_2(k_x,k_y)
\end{array} \right ],
\]
$t(k_z) = -2t_\perp\cos(k_zd/2)$, $d$ is the unit cell length along
the $c$ axis, and $c_{i{\bf k}}^\dagger$ creates an
electron in the sublattice $i$ with three dimensional momentum ${\bf k}$.
When the layers are coupled by $t_\perp$, they hybridize and form two bands 
with energies:
$ \epsilon_\pm = (\xi_1+\xi_2)/2 \pm \sqrt{(\xi_1-\xi_2)^2/2 + t(k_z)^2}$,
given by the eigenvalues of $h({\bf k})$.

The first necessary condition is that the layers be weakly coupled in
the sense that $t^2 \ll (\xi_1 - \xi_2)^2$ throughout most of the
Brillouin zone.  In this case the Fermi velocity along the $c$ axis is
$v_{z\pm}({\bf k}) \approx \pm {\partial t(k_z)}/{\partial k_z}
\times {t(k_z)}/{|\xi_1 - \xi_2|}$, which is of order $t_\perp^2$.  
In the usual expression for the $c$ axis conductivity, where one
neglects interband transitions,
\begin{equation}
\sigma_{zz}(\omega) 
= e^2 \sum_\pm \int \frac{d^3k}{4\pi^3} \,
\frac{v_{z\pm}({\bf k})^2}{2\Gamma-i\omega} 
\left (-\frac{\partial f}{\partial \epsilon_\pm} \right ),
\label{1a}
\end{equation}
where $f(x)$ is the Fermi function and $\Gamma = 1/2\tau$ is the 
scattering rate.  This expression is of order $t_\perp^4$.\cite{comment}

On the other hand, it is possible to show\cite{atkinsonI} that the
matrix element for an interband transition is $T_z \approx
\partial t(k_z)/\partial k_z$.  A simple estimate of the interband
conductivity from Fermi's golden rule is:
\[
\sigma_{\mbox{\scriptsize Inter}} = e^2\int \frac{d^3 k}{4\pi^2}
T_z^2 \, \frac{f(\epsilon_+)-f(\epsilon_-)}{\epsilon_+ - \epsilon_-} 
\delta(\omega-\epsilon_++\epsilon_-),
\]
which is obviously of order $t_\perp^2$.  

It is a general feature of these models, then, that one expects a crossover
from Drude-like (intraband dominated) to non-Drude-like (interband dominated)
conductivity along the $c$ axis as the strength of the interlayer coupling
is weakened.  

In order to proceed farther we choose $\xi_1 = k_\|^2/2m^\ast - \mu$
and $\xi_2 = \Delta_{\mbox{\scriptsize ps}}$, where $k_\|^2 =
k_x^2+k_y^2$ and $\Delta_{\mbox{\scriptsize ps}}$ is constant.  For
our crude calculation, we have taken $\xi_2$ to be a flat band (which
therefore has no Fermi surface!).  The basic features of the $c$ axis
conductivity are relatively insensitive to the choice of $\xi_2$.  The
scattering rate, $\Gamma(T)$, can be found by fitting the in-plane
resistivity.  For purposes of illustration, it is reasonable to take
$\Gamma(T) \equiv 1/2\tau = T$.

\begin{figure}[tb]
\begin{center}
\leavevmode
\epsfxsize .8\columnwidth
\epsffile{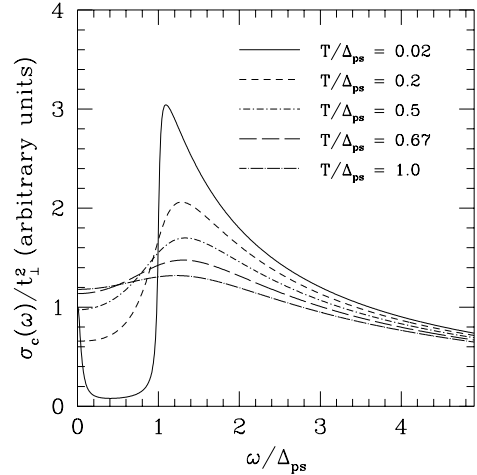}
\caption{The real part of the optical conductivity $\sigma_{c}(\omega)$
is shown for a range of temperatures.  The interplane
coupling is $t_\perp/\Delta_{\mbox{\scriptsize ps}} = 0.1$.  The in-plane
conductivity is given by the usual Drude expression, with corrections
of order $t_\perp^2/\Delta_{\mbox{\scriptsize ps}}^2$.}
\label{f1}
\end{center}
\end{figure}

We can now calculate the conductivity to linear order in the applied
field (see, e.g.~\onlinecite{atkinsonI}).  In the limit of weakly
coupled layers ($t_\perp \ll \Delta_{\mbox{\scriptsize ps}}$), the
real part of the optical conductivity is:
\begin{eqnarray}
\label{10}
\sigma_c(\omega) &=& 
{e^2 N_\| d t_\perp^2}
\Bigg[ \frac{4t_\perp^2}{\Delta_{\mbox{\scriptsize ps}}^2} \frac{2\Gamma}
{\omega^2+4\Gamma^2} \nonumber \\
&&+ 
\frac{1}{\omega} \left( \tan^{-1}
\left(\frac{\omega-\Delta_{\mbox{\scriptsize ps}}}
{\Gamma}\right)+ \tan^{-1}
\left(\frac{\omega+\Delta_{\mbox{\scriptsize ps}}}
{\Gamma}\right)\right) \nonumber \\
&&
\qquad \qquad \qquad \times \Theta(\mu+\Delta_{\mbox{\scriptsize ps}}-
\omega) \Bigg],
\end{eqnarray}
where $\Theta(x)$ is the step function.  In order to derive this
equation, we have made the approximation that the temperature
dependence of the conductivity is due mostly to the temperature
dependence of $\Gamma$.  Thermal excitations do not change the
conductivity in a qualitative fashion.  Equation~(\ref{10}) is plotted
in Fig.\ (\ref{f1}) for several different temperatures.

The first term in (\ref{10}) is the intraband conductivity and it has
the usual Drude form.  The second term is the interband conductivity.
True to our earlier assertion, we notice that the Drude term scales as
$t_\perp^2/\Delta_{\mbox{\scriptsize ps}}^2$ relative to the interband
term.

The interband conductivity has a number of features worth mentioning.
The first is that there is a threshold frequency $\omega =
\Delta_{\mbox{\scriptsize ps}}$ for interband transitions.
$\Delta_{\mbox{\scriptsize ps}}$ is the minimum energy for an
interband transition between a filled and an empty state.  This
threshold is smeared by electron scattering, and the gap in the
optical conductivity disappears at temperatures where $\Gamma(T) \sim
\Delta_{\mbox{\scriptsize ps}}$.  This kind of behaviour is seen in
optical conductivity experiments,\cite{homes} and it is tempting to
identify $\Delta_{\mbox{\scriptsize ps}}$ with the pseudogap seen
there.  If this is the case, then the relationship between the
``optical pseudogap'' and the pseudogap which opens in the Fermi
surface\cite{arpes} becomes clouded.  One expects that changes in the
Fermi surface will affect the interband transitions.  However, whether
or not the threshold energy (and therefore the optical pseudogap) is
affected will depend on the specifics of the band structure.  For this
reason, we draw a distinction between the optical pseudogap and the
pseudogap seen in other experiments.

While the band structure we have chosen here is too simple to describe
systems in which there is no optical pseudogap, we are at least able
to cite the conditions under which we expect the gap to disappear.
First, if the scattering rate $\Gamma(T)$ is always larger than the
minimum interband excitation energy, no pseudogap will be seen.  A
second possibility is that the minimum excitation energy between the
two bands may vanish at some $k$-point.  Such would be the case in YBCO, for
example, if the chain and plane Fermi surfaces were to cross\cite{atkinsonI}.

Another point worth noting is that the interband conductivity extends over
a broad range of frequencies, and is cut off by the band edge at
$\omega = \mu+\Delta_{\mbox{\scriptsize ps}}$.  With the
current model, the conductivity falls as $\omega^{-1}$.  The frequency
dependence is model specific, however, and it has been found that in
a model YBCO system, the conductivity can be quite flat.  In optimally
doped YBCO, interband transitions between the plane and chain bands allow
us to reconcile the broad $c$ axis spectrum (which seems to imply a large
scattering rate) with the relatively low value of the resistivity.

\begin{figure}[tb]
\begin{center}
\leavevmode
\epsfxsize .8\columnwidth
\epsffile{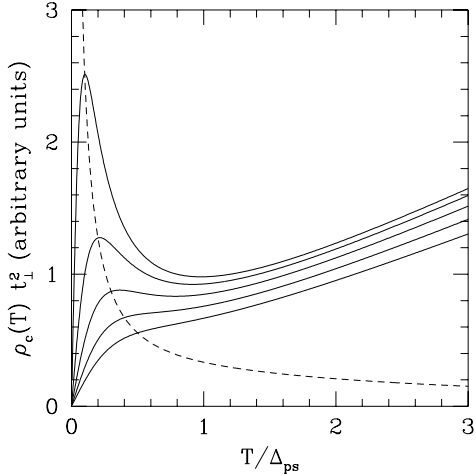}
\caption{The normalised d.c.\ resistivity $\rho_c(T) t_\perp^2$ 
is plotted as a function of temperature
for a range of interlayer coupling strengths $t_\perp$ between
$\Delta_{\mbox{\scriptsize ps}}/5$ and 
$\Delta_{\mbox{\scriptsize ps}}$ (solid curves, from top to bottom).
The Drude term in the resistivity is important only to the left of
the dashed curve.  The in-plane resistivity is given by usual Drude 
expression and increases linearly with $T$.}
\label{f2}
\end{center}
\end{figure}

The pseudogap also has an effect on the shape of the d.c.\ conductivity.
Taking the $\omega = 0$ limit of (\ref{10}) gives
\begin{equation}
\sigma_c(\omega=0) = 
{e^2 N_\| d t_\perp^2} 
\left[ \frac{4t_\perp^2}{\Delta_{\mbox{\scriptsize ps}}^2}\frac{1}{2\Gamma} 
+
\frac{2 \Gamma}{\Delta_{\mbox{\scriptsize ps}}^2+\Gamma^2} \right ],
\end{equation}
which is plotted in Fig.~\ref{f2}.
The first and second terms are again the intraband and interband
conductivities respectively.  If we consider the d.c.\ resistivity due
to the interband transitions alone, then
\[
 \rho_c \propto \left \{ \begin{array}{ll}
		1/\Gamma(T), & \Gamma(T) < \Delta_{\mbox{\scriptsize ps}} \\
		\Gamma(T), & \Gamma(T) > \Delta_{\mbox{\scriptsize ps}}
			\end{array}
		\right. .
\]
At very low temperatures, where $\Gamma$ is small, the system is short
circuited by the Drude conductivity.  Therefore, we have the additional low
temperature regime,
\[
  \rho_c \propto \Gamma(T),\quad \Gamma(T) < t_\perp.
\]
Of course, if there is some intrinsic disorder (in the chain layer in
YBCO, for example), then $\Gamma(T)$ may always be greater than $t_\perp$,
and the Drude term will never become apparent.  

Finally, we wish to examine the penetration depth.
In order to discuss the superconducting state, we make two changes to
the model.  First, we introduce a $d$ wave superconducting gap,
$\Delta_\phi = \Delta_0\cos(2\phi)$,
in each of the layers.  Second, we set the scattering rate
$\Gamma$ to zero.  There is good evidence that the scattering rate
does in fact drop dramatically below the superconducting transition
temperature $T_c$\cite{hardy2}, but the assumption is made primarily to 
simplify the model.

We have given elsewhere\cite{atkinsonII} an expression for the penetration
depth in a layered superconductor.  For this model it is possible to
find a simple low temperature expression for the penetration depth:
\begin{eqnarray}
\lambda_c(T)^{-2}& =& \lambda_c(0)^{-2}  \nonumber \\
&&-\frac{16\pi N_\|e^2}{c^2}\frac{t_\perp^2d}
{\Delta_{\mbox{\scriptsize ps}}^2}
\frac{T}{\Delta_0} \left[\ln(2) t_\perp^2 + 
\frac{3\zeta(3)}{2} T^2 \right ], \nonumber \\
\end{eqnarray}
where $\zeta$ is the Riemann zeta function.  Again, the first term in
the equation is due to the intraband conductivity and the second is
due to the interband conductivity.  Within a single band model, it is
usual to associate a linear temperature dependence with a $d$ wave
gap.  Here, the very low $T$ behaviour is linear in $T$, but there is
a crossover in behaviour to $T^3$ when $T \approx t_\perp.$ It is
interesting to note that the pseudogap affects both the interband and
the intraband terms in the same way, and that it is actually the
magnitude of $t_\perp$ which determines whether one sees a linear or
cubic temperature dependence.

This work was supported by the Natural Sciences and Engineering Research
Council, and the Canadian Institute for Advanced Research.  One of the
authors (W.\ A.) was supported in part by the Midwest
Superconductivity Consortium through D.O.E.
grant \# DE-FG-02-90ER45427.


\begin{thebibliography}{abrikosovvvv}
\bibitem{comment}  It is usual
to integrate Eq.~(\protect\ref{1a}) by parts in $k_z$, yielding the
familiar expression $\sigma_{zz}(\omega) = ne^2/m^\ast_{zz} /(2\Gamma
- i\omega)$.  This expression contains a factor of only $t_\perp^2$,
which is hidden in the effective mass, in apparent contradiction with
out assertion that $\sigma_{zz}$ scales as $t_\perp^4$.  It must be
remembered, however, that for an open Fermi surface, the integration
by parts yields an additional term which does not vanish.
\end{thebibliography}
\end{document}